\documentclass[12pt]{article}
\usepackage{amsmath,amssymb}
\usepackage{a4wide,epic,curves}

\title{
  The ``optical lever'' intracavity readout scheme for gravitational-wave antennae
}
\author{F.Ya.Khalili\footnote{farid@mol.phys.msu.su} \\
  {\it Dept. of Physics, Moscow State University}, \\
  {\it Moscow 199899, Russia}
}
\date{}

\begin{document}

\maketitle

\begin{abstract}

An improved version of the ``optical bar'' intracavity readout scheme for
gravitational-wave antennae is considered. We propose to call this scheme
``optical lever'' because it can provide significant gain in the signal
displacement of the local mirror similar to the gain which can be obtained
using ordinary mechanical lever with unequal arms. In this scheme
displacement of the local mirror can be close to the signal displacement of
the end mirrors of hypothetical gravitational-wave antenna with arm lengths
equal to the half-wavelength of the gravitational wave.

\end{abstract}

\section{Introduction}

All contemporary large-scale gravitational-wave antennae are based on common
principle: they convert phase shift of the optical pumping field into the
intensity modulation of the output light beam being registered by
photodetector \cite{Abramovici1992}. This principle allows to obtain
sensitivity necessary to detect gravitational waves from astrophysical
sources. However, its use in the next generations of gravitational-wave
antennae where substantially higher sensitivity is required, encounters
serious problems.

An excessively high value of optical pumping power which also depends sharply
on the required sensitivity, is likely to be the most important one. For
example, at the stage II of the LIGO project the light power circulating in
the interferometer arms will be increased to about 1 MWatt, in comparison
with about 10 KWatt being currently used \cite{WhitePaper1999}. In
particular, so high values of the optical power can produce undesirable
non-linear effects in the large-scale Fabry-Perot cavities
\cite{BSV_Instab2001}.

This dependence of pumping power on sensitivity can be explained easily using
the Heisenberg uncertainty relation. Really, in order to detect displacement
$\Delta x$ of test mass $M$ it is necessary to provide perturbation of its
momentum $ \Delta p \ge \hbar/2\Delta x$. The only source of this
perturbation in the interferometric gravitational-wave antennae is the
uncertainty of the optical pumping energy $\Delta\mathcal{E}$. Hence, the
following conditions have to be fulfilled: $\Delta\mathcal{E}\propto (\Delta
x)^{-1}$. If pumping field is in the coherent quantum state then
$\Delta\mathcal{E}\propto\sqrt\mathcal{E}$, and therefore $\mathcal{E}\propto
(\Delta x)^{-2}$.

Rigorous analysis (see \cite{Amaldi1999}) shows that pumping
energy stored in the interferometer have to be larger than
\begin{equation}\label{E_SQL}
  \mathcal{E} = \frac{ML^2\Omega^2\Delta\Omega}{4\omega_p\xi^2}\,,
\end{equation}
where $\Omega$ is the signal frequency, $\Delta\Omega<\Omega$ is
the bandwidth where necessary sensitivity is provided, $\omega_p$
is the pumping frequency, $L=c\tau$ is the length of the
interferometer arms, $\xi<1$ is the ratio of the amplitude of the
signal which can be detected to the amplitude corresponding to the
Standard Quantum Limit.

This problem can be alleviated by using optical pumping field in squeezed
quantum state \cite{Caves1981}, but can not be solved completely, because
only modest values of squeezing factor have been obtained experimentally yet.
Estimates show that usage of squeezed states allows to decrease $\xi$ by the
factor of $\simeq 3$ for the same value of the pumping energy (see
\cite{KLMTV2002}), and the energy still remains proportional to $\xi^{-2}$.

In the article \cite{NonLin1996} the new principle of {\em intracavity}
readout scheme for gravitational-wave antennae has been considered. It has
been proposed to register directly redistribution of the optical field {\em
inside} the optical cavities using Quantum Non-Demolition (QND) measurement
instead of monitoring output light beam.

The main advantage of such a measurement is that in this case a non-classical
optical field is created by the measurement process automatically. Therefore,
sensitivity of these schemes does not depend directly on the circulating
power and can be improved by increasing the precision of the intracavity
measurement device. The only fundamental limitation in this case is the
condition
\begin{equation}
  \frac{\Delta x}L \gtrsim \frac{\Omega}{\omega_p N}\,,
\end{equation}
where $N$ is the number of optical quanta in the antenna.

In the articles \cite{OptBar1997, SymPhot1998} two possible realizations of
this principle have been proposed and analyzed. Both of them are based on the
pondermotive QND measurement of the optical energy proposed in the article
\cite{JETP1977}. In these schemes displacement of the end mirrors of the
gravitational-wave antenna caused by the gravitational wave produces
redistribution of the optical energy between the two arms of the
interferometer. This redistribution, in its turn, produces variation of the
electromagnetic pressure on some additional local mirror (or mirrors). This
variation can be detected by measurement device which monitors position of
the local mirror(s) relative to reference mass placed outside the pumping
field (for example, a small-scale optical interferometric meter can be used
as such a meter).

The optical pumping field works here as a passive medium which transfers the
signal displacement of the end mirrors to the displacement of the local
one(s) and, at the same time, transfers perturbation of the local mirror(s)
due to measurement back to the end mirrors.

In this article we consider an improved version of the ``optical bar'' scheme
considered in the article \cite{OptBar1997}. We propose to call this scheme
``optical lever'' because it can provide gain in displacement of the local
mirror similar to the gain which can be obtained using ordinary mechanical
lever with unequal arms. This scheme is discussed in the section
\ref{sec:OptLeverL}.

In the section \ref{sec:OptLeverX} we analyse instability which can exist in
both ``optical bar'' and ``optical lever'' schemes (namely, in so-called
X-topologies of these schemes) and which was not mentioned in the article
\cite{OptBar1997}.

We suppose in this article for simplicity that all optical elements of the
scheme are ideal. It means that reflectivities of the end mirrors are equal
to unity, and all internal elements have no losses. We presume that optical
energy have been pumped into the interferometer using very small transparency
of some of the end mirrors, and at the time scale of the gravitation-wave
signal duration the scheme operates as a conservative one.

It has been shown in the article \cite{OptBar1997} that losses in
the optical elements limited the sensitivity at the level
\begin{equation}
  \xi \gtrsim \frac1{\sqrt{\Omega\tau_\mathrm{opt}^*}}
\end{equation}
where $\tau_\mathrm{opt}^*$ is the optical relaxation time. Taking into
account that value of $\tau_\mathrm{opt}^*$ can be as high as $\gtrsim
1\,\mathrm{s}$, one can conclude that the optical losses do not affect the
sensitivity if $\xi\gtrsim 10^{-1}$.

\section{The optical lever}\label{sec:OptLeverL}

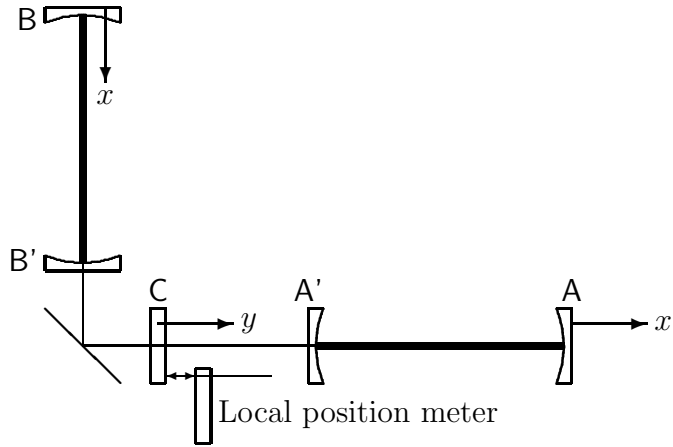
\begin{figure}

\begin{center}

\begin{picture}(100,60)

\thicklines

\thinlines\drawline(15,59)(15,15)(79,15)
\thicklines\drawline(46,14.75)(79,14.75)\drawline(46,15.25)(79,15.25)
\drawline(14.75,26)(14.75,59)\drawline(15.25,26)(15.25,59)

\drawline(78,10)(80,10)(80,20)(78,20)\curve(78,10,79,15,78,20)%
\put(80,21){\makebox(0,0)[cb]{\textsf{A}}}
\put(80,18){\vector(1,0){10}}\put(91,18){\makebox(0,0)[lc]{$x$}}%

\drawline(47,10)(45,10)(45,20)(47,20)\curve(47,10,46,15,47,20)%
\put(45,21){\makebox(0,0)[cb]{\textsf{A'}}}

\drawline(10,58)(10,60)(20,60)(20,58)\curve(10,58,15,59,20,58)%
\put(9,60){\makebox(0,0)[rt]{\textsf{B}}}%
\put(18,60){\vector(0,-1){10}}\put(18,49){\makebox(0,0)[ct]{$x$}}

\drawline(10,27)(10,25)(20,25)(20,27)\curve(10,27,15,26,20,27)%
\put(9,25){\makebox(0,0)[rb]{\textsf{B'}}}%

\drawline(10,20)(20,10)

\drawline(24,10)(24,20)(26,20)(26,10)(24,10)
\put(25,21){\makebox(0,0)[cb]{\textsf{C}}}
\put(25,18){\vector(1,0){10}}\put(36,18){\makebox(0,0)[lc]{$y$}}

\drawline(30,2)(30,12)(32,12)(32,2)(30,2)

\thinlines
\put(26,11){\vector(1,0){4}}\put(40,11){\vector(-1,0){14}}

\put(33,4){\makebox(0,0)[lb]{Local position meter}}

\end{picture}

\caption{L-topology of the ``optical lever'' scheme}\label{fig:OptLeverL}

\end{center}

\end{figure}

One of the possible ``optical lever'' scheme topologies (L-topology) is
presented in Fig.\,\ref{fig:OptLeverL} (another variant --- X-topology --- is
considered in the next section). It differs from the L-topology of the
``optical bar'' scheme \cite{OptBar1997} by two additional mirrors
\textsf{A'} and \textsf{B'} only. These mirrors together with the end mirrors
\textsf{A} and \textsf{B} form two Fabry-Perot cavities with the same initial
lengths $L=c\tau$ coupled by means of the central mirror \textsf{C} with
small transmittance $T_C$. Exactly as in the case of the ``optical bar''
scheme, due to this coupling eigen modes of such a system form the set of
doublets with frequencies separated by the value of $\Omega_B$ which is
proportional to the $T_C$ and can be made close to the signal frequency
$\Omega$.

Let distances between the mirrors to be adjusted in such a way that
Fabry-Perot cavities \textsf{AA'} and \textsf{BB'} are tuned in resonance
with the upper frequency of one of the doublets, and additional Fabry-Perot
cavities \textsf{A'C} and \textsf{B'C} are tuned in antiresonance with this
frequency.

It is supposed here that (i) distances $l$ between the mirrors \textsf{A',B'}
and the coupling mirror \textsf{C} are small enough to neglect values of the
order of magnitude close to $\Omega l/c$, and that (ii) it is possible to
neglect the mirrors \textsf{A',B'} motion. For example, they can be attached
rigidly to the platform where the local position meter is situated.

Let only the upper frequency of this doublet to be excited initially. In this
case most of the optical energy is concentrated in the cavities \textsf{AA'}
and \textsf{BB'} and distributed evenly between them. Small differential
change $x$ in the cavities optical lengths will redistribute the optical
energy between the arms and hence will create difference in pondermotive
forces acting on the mirrors. In other words, an optical pondermotive
rigidity exists in such a scheme.

In the article \cite{OptBar1997} the analogy with two ``optical springs'',
one of which was situated between the mirrors \textsf{A} and \textsf{C} and
another one (L-shaped) between the mirrors \textsf{B} and \textsf{C}, has
been used. It has been shown in that article that if the optical energy
exceeded the threshold value of
\begin{equation}\label{E_OB}
  \mathcal{E} \sim \frac{ML^2\Omega^3}{\omega_p} \,,
\end{equation}
then these springs became rigid enough to transfer displacement
$x$ of the mirrors \textsf{A}, \textsf{B} to the same displacement
$y$ of the mirror \textsf{C}.

It is rather evident that if additional mirrors \textsf{A',B'} are present
then displacement $x$ of the end mirrors \textsf{A,B} is equal to about
$\mathcal{F}$ times greater displacement $x$ of the mirrors \textsf{C}, where
$\mathcal{F}$ is finesse of the Fabry-Perot cavities \textsf{AA'} and
\textsf{BB'} (one can imagine, for simplicity, that these Fabry-Perot
cavities are replaced by delay lines). Therefore, one can expect that scheme
presented in Fig.\,\ref{fig:OptLeverL} provides gain in the mirror \textsf{C}
displacement relative to its displacement in the original ``optical bars''
scheme, and this gain has to be close to $\mathcal{F}$.

The analysis shows that it is true. The mechanical degrees of freedom
equations of motion in spectral representation have the form (we omit here
some very lengthy but rather straightforward calculations devoted to
excluding variables for electromagnetic degrees of freedom and reducing the
full equations set for the system to mechanical equations only):
\begin{subequations}\label{MechEq}
  \begin{align}
    \left[-2M_x\Omega^2 + K_{xx}(\Omega)\right]x(\Omega) &=
      K_{xy}(\Omega)y(\Omega) + F_\mathrm{grav}(\Omega) \,, \\
    \left[-M_y\Omega^2 + K_{yy}(\Omega)\right]y(\Omega) &=
      K_{xy}(\Omega)x(\Omega) + F_\mathrm{fluct}(\Omega) \,,
  \end{align}
\end{subequations}
Here $M_x$ is the mass of the mirrors \textsf{A,B}, $M_y$ is the mass of the
mirror \textsf{C}, $x$ is the displacement of the mirrors \textsf{A,B}, $y$
is the displacement of the mirror \textsf{C}, $F_\mathrm{grav}(\Omega)$ is
the signal force acting on the mirrors \textsf{A,B}, and $F_\mathrm{fluct}$
is fluctuational back action force produced by the device which monitors
variable $y$. We consider here only differential motion of the mirrors
\textsf{A,B}, when their displacements have the same absolute value and the
directions shown in Fig.\,\ref{fig:OptLeverL}. This motion corresponds to the
gravitational wave with optimal polarization. It has been shown in the
article \cite{OptBar1997} that the symmetric motion of the mirrors
\textsf{A,B} (when both mirrors approach to or move from the mirror
\textsf{C}) did not coupled with the degrees of freedom $x$ and $y$ and could
be excluded from the consideration.

Factors $K_{xx},K_{yy},K_{xy},K_{yx}$ which form the matrix of the
pondermotive rigidities are equal to
\begin{subequations}\label{K_OptLeverL}
  \begin{align}
    K_{xx}(\Omega) &= \frac{2\omega_p\mathcal{E}}{c^2\tau\cos^2\Omega\tau}\,
      \frac{\tan\Omega_B\tau}{\tan^2\Omega_B\tau-\tan^2\Omega\tau} \,, \\
    K_{yy}(\Omega) &= \frac{2\omega_p\mathcal{E}}{c^2\tau\digamma^2}\,
      \frac{\tan\Omega_B\tau}{\tan^2\Omega_B\tau-\tan^2\Omega\tau} \,, \\
    K_{xy}(\Omega) &=\frac{2\omega_p\mathcal{E}}
      {c^2\tau\digamma\cos\Omega\tau} \,
      \frac{\tan\Omega_B\tau}{\tan^2\Omega_B\tau-\tan^2\Omega\tau}
      \,,
  \end{align}
\end{subequations}
where
\begin{equation}\label{R}
  \digamma = \frac{1+R}{1-R} \approx \frac{2}{\pi}\,\mathcal{F} \,,
\end{equation}
\begin{equation}\label{Omega_B}
  \Omega_B = \frac1\tau\arctan\left(\frac{\tan\phi}{\digamma}\right)\,,
\end{equation}
$\phi=\arcsin T_C$ and $R$ is the reflectivity of the mirrors \textsf{A',B'}.
It have to be noted that these rigidities exactly satisfy the condition
\begin{equation}
  K_{xx}(\Omega)K_{yy}(\Omega) - K_{xy}^2(\Omega) = 0 \,,
\end{equation}

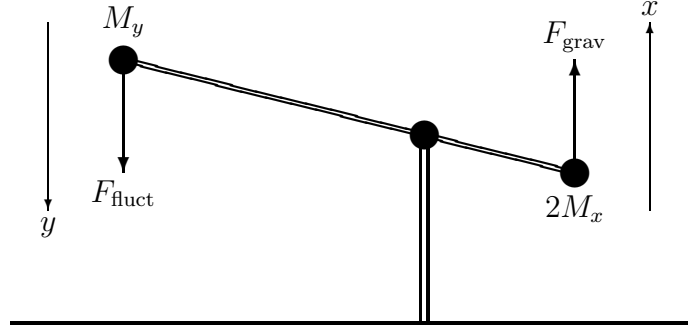
\begin{figure}

\begin{center}

\begin{picture}(90,45)

\thinlines

\drawline(0,0)(90,0)

\put(5,40){\vector(0,-1){25}}\put(5,14){\makebox(0,0)[ct]{$y$}}

\put(85,15){\vector(0,1){25}}\put(85,41){\makebox(0,0)[cb]{$x$}}

\thicklines

\drawline(0,0)(90,0)\drawline(54.5,0)(54.5,25)(55.5,25)(55.5,0)%
\put(55,25){\circle*{4}}

\drawline(16,35)(76,20)\drawline(14,35)(74,20)

\put(15,35){\circle*{4}}\put(15,38){\makebox(0,0)[cb]{$M_y$}}%
\put(15,35){\vector(0,-1){15}}%
\put(15,19){\makebox(0,0)[ct]{$F_\mathrm{fluct}$}}

\put(75,20){\circle*{4}}\put(75,17){\makebox(0,0)[ct]{$2M_x$}}%
\put(75,20){\vector(0,1){15}}%
\put(75,36){\makebox(0,0)[cb]{$F_\mathrm{grav}$}}

\end{picture}

\caption{Mechanical model of the ``optical lever'' scheme}\label{Fig:Lever}

\end{center}

\end{figure}

Suppose then $\Omega\tau \ll 1$ (in the case of the contemporary terrestrial
gravitational-wave antennae $\tau\lesssim 10^{-5}\,\mathrm{s}$ and
$\Omega\lesssim 10^3\,\mathrm{s}^{-1}$, so $\Omega\tau \lesssim 10^{-2})$. In
this case we obtain that
\begin{equation}\label{K_xx_short}
  K_{xx}(\Omega) = \frac{2\omega_p\mathcal{E}}{L^2}\,
    \frac{\Omega_B}{\Omega_B^2-\Omega^2} \,,
\end{equation}
and
\begin{equation}\label{K_lever}
  K_{xx}(\Omega) = \digamma K_{xy} = \digamma^2 K_{yy} \,.
\end{equation}
There is a very simple mechanical model which also can be described by the
equations (\ref{MechEq},\ref{K_lever}), putting aside for a while particular
spectral dependence (\ref{K_xx_short}). This is an ordinary mechanical lever
with arm lengths ratio $\digamma$ (see Fig.\,\ref{Fig:Lever}). Rigidities
$K_{xx}, K_{yy}$, and $K_{xy}$ in this case are proportional to the bending
rigidity of the lever bar. It is evident that if the motion is sufficiently
slow and therefore it is possible to neglect bending then the $y$-arm tip
will be $\digamma$ times greater than the $x$-arm tip displacement.
Consequently, if the observation frequency $\Omega$ is sufficiently small
then in the ``optical lever'' scheme presented in Fig.\,\ref{fig:OptLeverL}
the mirror \textsf{C} motion will repeat the end mirrors \textsf{A,B} motion
with the gain factor $\digamma$.

In all other aspects this scheme is similar to the ``optical bars'' scheme.
As it follows from the symmetry conditions (\ref{K_lever}) if one replaces in
the equations (\ref{MechEq}) $y$ by $y/\digamma$, $F_\mathrm{fluct}$ by
$\digamma\times F_\mathrm{fluct}$, $M_y$ by $\digamma^2\times M_y$ and then
replaces all rigidities by their values corresponding to ``optical bars''
scheme (with $\digamma=1$) then these equations still remain valid.

It means that if in the ``optical bars'' scheme one (i) replaces the mass
$M_y$ by $\digamma^2$ smaller one; (ii) decreases back action noise of the
meter by the factor of $\digamma$ and increases proportionally its
measurement noise (for example, by decreasing pumping power in the
interferometric position meter by the factor of $\digamma^2$); and (iii)
inserts the additional mirrors \textsf{A',B'} with refelctivity defined by
the equation (\ref{R}) then signal-to-noise ratio (relative to the local
meter noises) and dynamical properties of the scheme will remain unchanged,
with the only evident replacement of $y$ by $\digamma y$.

Two characteristic regimes of the ``optical lever'' scheme are similar to the
quasistatic and resonant regimes of the ``optical bars'' scheme described in
the article \cite{OptBar1997}, and therefore here we consider them in brief
only.

Characteristic equation of the equations set (\ref{MechEq}) is the following:
\begin{equation}\label{CharactEq}
  \Omega^2\left(\Omega^4 - \Omega_B^2\Omega^2
    + \frac{2\omega_p\mathcal{E}\Omega_B}{M_\mathrm{eff}L^2}\right) = 0 \,,
\end{equation}
where
\begin{equation}
  M_\mathrm{eff} = \left(\frac1{2M_x} + \frac1{\digamma^2M_y}\right)^{-1} \,.
\end{equation}
Root $\Omega=0$ of this equation corresponds to the quasistatic regime. If
pumping energy is sufficiently high:
\begin{align}\label{CondForE}
  K_{xx}(\Omega) &= \frac{2\omega_p\mathcal{E}}{L^2}\,
    \frac{\Omega_B}{\Omega_B^2-\Omega^2} \gg M_x\Omega^2 \,,     &
  K_{yy}(\Omega) &= \frac{2\omega_p\mathcal{E}}{\digamma^2 L^2}\,
    \frac{\Omega_B}{\Omega_B^2-\Omega^2} \gg M_y\Omega^2 \,,
\end{align}
then the equations set (\ref{MechEq}) solution for this regime can
be presented as
\begin{equation}
  y(\Omega) = \digamma x(\Omega) = -\frac{
    \digamma F_\mathrm{grav}(\Omega) + \digamma^2F_\mathrm{fluct}(\Omega)
  }{\Omega^2(2M_x + \digamma^2 M_y)} \,.
\end{equation}
It is evident that maximal value of the signal response can be
obtained here if
\begin{equation}\label{Masses}
  \digamma = \sqrt{\frac{2M_x}{M_y}} \,,
\end{equation}
and in this case we will get
\begin{equation}
  y(\Omega) = \digamma x(\Omega) = -\frac1{2\Omega^2}\left(
    \frac{F_\mathrm{grav}(\Omega)}{\sqrt{2M_xM_y}}
    + \frac{F_\mathrm{fluct}(\Omega)}{M_y}
  \right) \,.
\end{equation}
Taking into account, that gravitational-wave signal force is proportional to
the mass of the end mirrors: $F_\mathrm{grav}\propto M_x$, we can conclude
that in the gravitational-wave experiments this regime can provide a
wide-band gain in signal displacement proportional to $\sqrt{M_x/M_y}$.

It is necessary to note that this gain by itself does not allow to overcome
the standard quantum limit because the value of the standard quantum limit
for the test mass $M_y$ rises exactly in the same proportion. But it does
allow to use less sensitive local position meter and it does increase the
signal-to-noise ratio for miscellaneous noises of non-quantum origin and
therefore makes it easier to overcome the standard quantum limit using, for
example, variational measurement in the local position meter
\cite{Vyatchanin1998,DSVM2000}.

Another two roots of the equation (\ref{CharactEq})
\begin{equation}
  \Omega_{1,2}^2 = \frac{\Omega_B^2}2 \pm \sqrt{
    \frac{\Omega_B^2}2
    - \frac{4\omega_p\mathcal{E}\Omega_B}{M_\mathrm{eff}L^2}
  }
\end{equation}
correspond to the more sophisticated resonant regime of the scheme. Placing
these two roots evenly in the spectral band of the signal it is possible to
obtain sensitivity a few times better than the standard quantum limit for a
free mass in relatively wide band, as it has been proposed in the article
\cite{Buonanno2001}. Using the value of pumping power
\begin{equation}
  \mathcal{E} = \frac{M_\mathrm{eff}L^2\Omega_B^3}{8\omega_p} \,,
\end{equation}
it is possible to implement the second-order-pole test object and obtain the
sensitivity substantially better than the standard quantum limit for both
free mass and harmonic oscillator in narrow band near the frequency
$\Omega_B/\sqrt2$ \cite{FDRigidity2001}. In both cases ``optical lever''
allows to increase the signal displacement of the local mirrors and therefore
makes it easier implementation of the local position meter.

\section{X-topologies of the ``optical bars'' and the
``optical lever'' schemes}\label{sec:OptLeverX}

\begin{figure}

\begin{center}

\begin{picture}(90,70)

\thicklines
\drawline(30.5,19.5)(29.5,20.5)(39.5,30.5)(40.5,29.5)(30.5,19.5)
\put(29,19){\makebox(0,0)[rt]{\textsf{C}}}%

\thicklines       
\drawline(78,20)(80,20)(80,30)(78,30)\curve(78,20,79,25,78,30)%
\put(80,19){\makebox(0,0)[ct]{\textsf{A}}}%
\put(80,28){\vector(1,0){10}}\put(90,27){\makebox(0,0)[rt]{$x$}}%

\drawline(45,20)(43,20)(43,30)(45,30)\curve(45,20,44,25,45,30)%
\put(45,19){\makebox(0,0)[ct]{\textsf{A'}}}%

\thicklines       
\drawline(30,68)(30,70)(40,70)(40,68)\curve(30,68,35,69,40,68)%
\put(29,70){\makebox(0,0)[rt]{\textsf{B}}}%
\put(38,70){\vector(0,-1){10}}\put(38,59){\makebox(0,0)[ct]{$x$}}

\drawline(30,35)(30,33)(40,33)(40,35)\curve(30,35,35,34,40,35)%
\put(30,35){\makebox(0,0)[rb]{\textsf{B'}}}%

\thicklines       
\drawline(12,20)(10,20)(10,30)(12,30)\curve(12,20,11,25,12,30)%
\put(10,19){\makebox(0,0)[ct]{\textsf{B''}}}%
\put(10,28){\vector(-1,0){10}}\put(0,27){\makebox(0,0)[lt]{$y$}}

\thicklines       
\drawline(30,2)(30,0)(40,0)(40,2)\curve(30,2,35,1,40,2)%
\put(29,0){\makebox(0,0)[rb]{\textsf{A''}}}%
\put(38,0){\vector(0,1){10}}\put(38,10){\makebox(0,0)[cb]{$y$}}

\thinlines\drawline(10,25)(80,25)\drawline(35,0)(35,70)
\thicklines\drawline(44,24.75)(79,24.75)\drawline(44,25.25)(79,25.25)
\drawline(34.75,34)(34.75,69)\drawline(35.25,34)(35.25,69)

\end{picture}

\caption{X-topology of the ``optical lever'' scheme}\label{Fig:OptLeverX}

\end{center}

\end{figure}
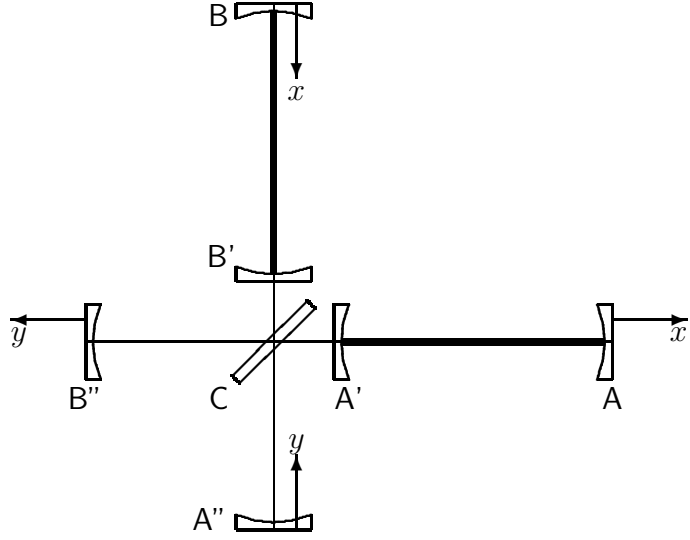

In the article \cite{OptBar1997} two possible topologies of the ``optical
bars'' scheme have been considered, the L-topology discussed in the previous
section and the X-topology similar to the Michelson interferometer topology.
The latter one also can be converted to the ``optical lever'' scheme using
additional mirrors \textsf{A'} and \textsf{B'} as it is shown in
Fig.\,\ref{Fig:OptLeverX}. Here \textsf{C} is the coupling mirror with
transmittance $T_C=\sin\phi$. In this topology one optical ``spring'' exists
between the mirrors \textsf{A} and \textsf{A''}, and another one between the
mirrors \textsf{B} and \textsf{B''}

In the article \cite{OptBar1997} L- and X-topologies were considered as
identical ones with the only difference that in the case of X-topology the
value of $\Omega_B$  was about two times greater,
\begin{equation}
  \Omega_B = \frac1\tau\arctan\left(\frac{\tan 2\phi}{\digamma}\right)\,.
\end{equation}
More rigorous analysis shows, however, that this is not the case. Really, the
rigidities which appear in the equations (\ref{MechEq}) for the case of the
X-topology are equal to
\begin{subequations}\label{K_OptLeverX}
  \begin{align}
    K_{xx}(\Omega) &= \frac{2\omega_p\mathcal{E}}{c^2\tau\cos^2\Omega\tau}\,
      \frac{\tan\Omega_B\tau}{\tan^2\Omega_B\tau-\tan^2\Omega\tau} \,, \\
    K_{yy}(\Omega)
      &=\frac{2\omega_p\mathcal{E}(1+\digamma^2\tan^2\Omega\tau)}
        {c^2\tau\digamma^2}\,
        \frac{\tan\Omega_B\tau}{\tan^2\Omega_B\tau-\tan^2\Omega\tau} \,, \\
    K_{xy}(\Omega) &=\frac{2\omega_p\mathcal{E}}
      {c^2\tau\digamma\cos\Omega\tau\cos 2\phi} \,
      \frac{\tan\Omega_B\tau}{\tan^2\Omega_B\tau-\tan^2\Omega\tau}
      \,,
  \end{align}
\end{subequations}
and
\begin{equation}
  K_{xx}(\Omega)K_{yy}(\Omega)-K_{xy}^2(\Omega) =
  - \left(\frac{2\omega_p\mathcal{E}}{c^2\tau\cos\Omega\tau}\right)^2\,
      \frac{\tan^2\Omega_B\tau}{\tan^2\Omega_B\tau-\tan^2\Omega\tau}\ne 0 \,.
\end{equation}
It means that the low-frequency mechanical mode which in the article
\cite{OptBar1997} has been considered as a free mass mode (see factor $p^2$
in the equation (C.3) in the above-mentioned article) and which does
represent free mass in the case of L-topology, has a non-zero rigidity in the
case of X-topology. Moreover, if $\Omega<\Omega_B$ then this rigidity is
negative, and therefore asynchronous instability exists in the
system\footnote{In the article \cite{OptBar1997} another instability has been
considered which has different origin and depends on the optical relaxation
time. Instability we consider here exists even if this relaxation time is
equal to infinity.}.

Characteristic time for this instability is equal to
\begin{equation}
  \tau_\mathrm{instab} = \left(
    \frac{2\omega_p\mathcal{E}\Omega_B}{L^2(2M_x/\digamma^2+M_y)}
  \right)^{-1/2} \,.
\end{equation}
Taking into account condition (\ref{CondForE}) an supposing that
$\Omega\sim\Omega_B$, one can obtain that
\begin{equation}
  \tau_\mathrm{instab}\Omega \approx \frac1{\digamma\Omega\tau} \,.
\end{equation}
This value is rather large ($\sim 10^2$ if, say, $\Omega\sim
10^3\,\mathrm{s}^{-1}$ and $\tau\sim 10^{-5}\,\mathrm{s}$) in the case of
pure ``optical bars'' scheme ($\digamma=1$). Therefore, this instability can
be easily dumped by the feed-back system in this case. On the other hand, in
the case of the ``optical lever'' scheme can be $\tau_\mathrm{instab} \sim
\Omega^{-1}$, if one attempts to use too large value of $\digamma$.

In the article \cite{Buonanno2002} it has been shown, however, that even such
a strong instability can be dumped in principle by feed-back scheme without
any loss in the signal-to-noise ratio.

\section{Conclusion}

Properties of the ``optical bars'' intracavity scheme
\cite{OptBar1997} can be substantially improved by converting arms
of the antenna into Fabry-Perot cavities similar to ones used in
traditional topologies of gravitational-wave antennae with
extracavity measurement. This new ``optical lever'' scheme allows
to obtain the gain in signal displacement of local mirror
approximately equal to finesse of the Fabry-Perot cavities.

This gain by itself does not allow to overcome the standard
quantum limit in wide-band regime. But it allows to use less
sensitive local position meter and increases the signal-to-noise
ratio for miscellaneous noises of non-quantum origin, making it
easier to overcome the standard quantum limit using, for example,
variational measurement in the local position meter.

The value of this gain is limited, in principle, by the formula
(\ref{Omega_B}) only. As it follows from this formula $\digamma$ can not
exceed value $(\Omega_B\tau)^{-1}$. If $\Omega_B \sim \Omega \sim
10^3\,\text{s}^{-1}$ and $\tau \sim 10^{-5}\,\text{s}$ (which corresponds to
arms length of LIGO and VIRGO antennae) then $\digamma \lesssim 10^2$. If
$\tau\sim 10^{-6}\,\text{s}$ (GEO-600 and TAMA) then this limitation is about
one order of magnitude less strong, $\digamma \lesssim 10^3$\,. It is
interesting to note that if $\digamma$ is close to its limiting value
$(\Omega_B\tau)^{-1} \sim (\Omega\tau)^{-1}$ then signal displacement of the
local mirror is close to the signal displacement of the end mirrors of
hypothetical gravitational-wave antenna with arm lengths equal to the
half-wavelength of the gravitational wave.

\section*{Acknowledgments}

Author thanks V.B.Braginsky, M.L.Gorodetsky, Yu.I.Vorontsov and
S.P.Vyatchanin for useful remarks and discussions.

This paper was supported in part by the California Institute of Technology,
US National Science Foundation, by the Russian Foundation for Basic Research,
and by the Russian Ministry of Industry and Science.

\end{document}